\def\etal{{\it et al.~\/}}

\def\ie{{\it i.e.~\/}}
\def\eg{{\it e.g.~\/}}
\def\ltsima{$\; \buildrel < \over \sim \;$}
\def\simlt{\lower.5ex\hbox{\ltsima}}
\def\gtsima{$\; \buildrel > \over \sim \;$}
\def\simgt{\lower.5ex\hbox{\gtsima}}
\def\att{${\cal A}_\lambda$~}

\documentstyle[12pt,aasms4,psfig]{article}
\tighten
\singlespace

\lefthead{A. Ferrara}
\righthead{}

\begin{document}

\title{An Atlas of Monte Carlo Models of Dust Extinction in Galaxies \\
for Cosmological Applications}

\author{Andrea Ferrara$^1$, Simone Bianchi$^2$, Andrea Cimatti$^1$ and
Carlo Giovanardi$^1$}
\affil{
$^1$Osservatorio Astrofisico di Arcetri \\ 50125 Firenze, Italy 
\\ E--mail: ferrara@arcetri.astro.it\\
E--mail: cimatti@arcetri.astro.it\\
E--mail: giova@arcetri.astro.it\\
$^2$Dept. of Physics and Astronomy, University of Wales Cardiff\\
P.O. Box 913, Cardiff Wales, CF2 3YB,  UK
\\ E--mail: Simone.Bianchi@astro.cf.ac.uk\\} 
\begin{abstract}
We present an extensive study of the radiative transfer in dusty galaxies
based on Monte Carlo simulations. 
The main output of these simulations are the   
attenuation curves ${\cal A}_\lambda$ (\ie the ratio between the observed, 
dust extinguished, total intensity to the intrinsic unextinguished one of the
galaxy as a function of wavelength). We have explored the dependence of
${\cal A}_\lambda$ on a conspicuous set of quantities
(Hubble type, inclination, dust optical thickness, dust distribution and
extinction properties) for a large wavelength interval, ranging from 
1250\AA~to the K band, thus finally providing a comprehensive atlas of
dust extinction in galaxies, which is electronically available. 
This study is particularly suitable for inclusion into galaxy formation
evolution models and to directly interpret observational data on high
redshift galaxies. 
\end{abstract}


\section{Motivation}

Modelling galaxy formation at high redshift, both via semi-analytical
methods and numerical simulations, has become 
one of the most active areas in cosmology in the last decade. 
As the observational identification  of galaxies is proceeding very rapidly to high
redshift (the present record being held by a galaxy at redshift $z=5.60$,
Spinrad \etal 1998)  thanks to the well-established Lyman-dropout 
technique (Steidel \etal 1996, Madau \etal 1996), a considerable effort
has to be devoted to the development of models of increasing complexity,
necessary to interpret the observed properties of such objects.
The detection of distant galaxies made possible by recent UV/visible surveys 
has allowed investigators  to tackle the problem  of the global star formation
rate in the universe (SFR) in a quantitative manner. We have now clarified that 
stars were forming at $z \approx 1$ at a rate about ten times higher than at 
present. During this period, the forming galaxies have converted a considerable
fraction of baryons into stars, an effect that is it likely seen in the decrease of 
the cold gas comoving density between $z=2$ and $z=0$. The evolution 
of the SFR at even higher redshift  is, however, not yet fully established.
The main difficulty  to this concern is the fact that approaches based on the 
dropout technique are poorly sensitive to dust-obscured galaxies.  
Hence, the SFR deduced in this way could represent a severe
underestimate of the actual one, if even a rather modest amount of dust is
present in the interstellar medium of the star forming galaxy. Also, some 
galaxies could be so heavily extinguished that they could be completely 
missed from the UV/visible census.  Support to this possibility is lent by
available IR (Rowan-Robinson \etal 1997) and sub-mm (Smail \etal 1997, Cimatti
\etal 1998) 
observations that have explicitly tackled this problem. Using the modelled
SED of 12 galaxies detected by ISO in the HDF, Smail \etal derived 
star formation rates significantly higher that those derived by Madau \etal 
(1996). From the analysis of source counts 
at two wavelengths (450 and 850 $\mu$m) in a survey exploiting the
high sensitivity SCUBA bolometer array, these authors also concluded 
that the number density of
star forming galaxies strongly increases in the high redshift universe. 
These results are further substantiated by the similar conclusions 
reached by Franceschini \etal 1997, who studied K-band selected 
early type galaxies in the HDF, and by statistical approaches as those
in Pozzetti \etal 1998.

At present, probably the most successful class of models developed 
to study the formation and evolution of galaxies are based on the 
hierarchical paradigm, \ie structure in the universe grows by 
association of small units to form larger bound objects.
Starting from a number of (more or less heuristic) prescriptions ruling 
the gas collapse in the dark matter halos, the process of star formation, 
stellar feedback and evolution, these models are able to predict several 
observables
characterizing real galaxies. This approach has been refined to an
increasing degree of sophistication and applied to different 
aspects by a large number of groups (White \& Frenk 1991, Lacey \etal
1993, Kauffman 1995, Ciardi \& Ferrara 1997, Baugh \etal 1997,
Guiderdoni \etal 1998). However, at least one aspect of this type of studies
-- dust effects -- has probably been overlooked or at best only very 
coarsely treated. An effort is therefore required to include a 
proper dust treatment not only in these semi-analytical type of models,
but also in extensive numerical simulations and even in more direct data 
interpretations.
Typically,  such studies make very simple (and sometimes crude) 
approximations about the dust component of galaxies. A standard set 
would, for example, assume that (i) the dust is distributed as the stars,
(ii) inclination-dependent extinction effects can be averaged, (iii) the
scattering of grains is often assumed as isotropic and the albedo 
treated in a heuristic way, and finally (iv) "screen" or "sandwich"  
geometries are adopted, in which there is a layer of pure dust between the
observer and any stars. In addition, physical effects related to the 
different optical thickness of protogalaxies (which according to the
above discussion could be substantial) and to the possibly 
different extinction curve remain largely to be explored. 
The assumption of "screen" geometries, resulting in a more efficient
extinction of galactic light with respect to distributions with the dust
and stars mixed,
typically tends to underestimate the amount of dust. This effect is 
confirmed by comparison with FIR/sub-mm observations, which are
sensitive to dust thermal emission, which often yield dust masses
larger than expected on the basis of screen geometry absorption estimates. 
Using realistic dust distributions and detailed radiative transfer
helps reconcile the above discrepancy. For example, our mass estimate
for the red galaxy HR10 (Cimatti \etal 1997) using Monte Carlo
simulations of the type presented here is in close agreement 
with the one obtained from sub-mm data (Cimatti \etal 1998). 
The ongoing work by Cole \etal (1998),
which is using some of the present results, and the recent study by
Silva \etal (1998) modelling the effects of dust on galactic SEDs,
very likely are setting the scene for additional progress in this area. 

Only recently more realistic models have been developed (Kylafis \& 
Bahcall 1987; Bruzual et al.  1988; Witt et al. 1992; Byun et al. 1994; 
Wise \& Silva 1996; Bianchi, Ferrara \& Giovanardi 1996, hereafter BFG). 
Although these models have shown that dust
scattering plays an important role by reducing the effects of dust
absorption, no systematic studies of the effects of the extinction on the
observed colours of high-$z$ galaxies have been performed. Understanding
these effects is crucial in cosmology and galaxy evolution studies.

The main aim of this work is to provide a large database of dust 
extinction cases that could be safely included into galaxy 
evolution models, numerical simulations, and interpretation of
observational results. The way in which we accomplish this task
is by using the Monte Carlo numerical code originally developed
to simulate the extinction and polarization properties of spiral
galaxies (described in Bianchi, Ferrara \& Giovanardi 1996).
The code has been adapted  to include a number of aspects
discussed below, and to obtain handy and directly relevant quantities
as the attenuation curves ${\cal A}_\lambda$ (\ie the ratio between the observed, dust
extinguished, total intensity to the intrinsic unextinguished one of the
galaxy as a function of wavelength). We have explored the dependence of 
${\cal A}_\lambda$ on a set of observable quantities
(Hubble type, inclination, dust optical thickness, dust distribution and
extinction properties) for a large wavelength interval, ranging from the 
1250\AA to the K band, thus finally providing a comprehensive atlas of
dust extinction in galaxies. This study could be used in the future for
a number of different applications; as an example, Cimatti \etal (1997)
have exploited some of the results presented here to investigate the
problem of the age-dust degeneracy in high-$z$ elliptical galaxies.

The plan of the paper is the following. In \S~2 we present in detail 
our dusty galaxy models, emphasizing the changes with respect
to BFG, and the model classification adopted;  \S~3 is devoted to
the results, mainly constituted by the attenuation curves for the
various cases (obviously, due to the huge amount of data only 
a small subset are shown here, but attenuation curves for 
all the studied cases are available in electronic form -- see \S~5).
Finally, \S~4 contains a few additional comments concerning the
general applicability of our results. 

\section{Dusty Galaxy Models}

In order to study the effects of dust extinction,  a realistic model
for the light and dust distribution in galaxies is required.
We start by dividing galaxies in two broad classes according to
their stellar distribution: spirals  and  ellipticals,
respectively. The former are characterized by a disk and a bulge,
whereas ellipticals consist of a bulge only. 

For both types of galaxies 
we consider two different dust extinction curves, reproducing
the results for the (i) Milky Way (MW) and (ii) the
Small Magellanic Cloud (SMC) (Gordon \etal 1997). 
The adopted curves are shown in Fig. \ref{fig1}. The main difference between 
them is the relevance of the 2175\AA\ bump, absent in the SMC extinction 
law. The physical interpretation of such a feature in terms of the
type, shape and eventually coating of the grains producing it
remains still unclear (for a discussion see, for example, Draine \& 
Malhotra 1993). The     common wisdom is to assume that in starburst
galaxies the bump is weak or absent. This idea is based on 
observations of the extinction curve in the 30 Dor region,  
(often considered as a starburst prototype)
showing that the feature is very weak, whereas outside the
region the extinction curve is similar to the Galactic one
(Fitzpatrick 1985, 1986). The bump has been recently detected
also in high redshift Mg~II absorbers (Malhotra 1997), thus
suggesting that the dust properties in the young universe 
could be rather similar to the local ones.
Finally, one has to keep in mind that the absence of this
extinction feature in external galaxies could be due 
either to a different dust composition or to scattering effects
compensating for dust absorption, as shown by Cimatti \etal (1997). 

We explore several possible distributions of galactic  dust in order to 
cover a wide range of possible applications, which are described in detail 
below. For each distribution we have studied 9 values of
the optical depth in the V-band ($\tau_V=0, 0.1, 0.5, 1, 2, 5, 10, 20, 
50$) in the rest frame of the galaxy along a line of sight passing through 
the center of the galaxy (perpendicular to the galactic plane, when a dust
disk is present). The total galactic dust mass can be 
calculated from  a given value of $\tau_V$ by using the conversion
formulae given in Appendix A, which depend on grain properties and dust 
distribution. 

\subsection{Spiral Galaxies}

\subsubsection{Stellar Distribution}

Following BFG, we describe the stellar spatial distribution in spirals by 
two components:
a spheroidal bulge and a three-dimensional disk. We neglect dark, massive
halos, since we are interested only in luminous components, and possible
small scale inhomogeneities, such as spiral arms and/or star-forming
regions.  

\underline{Disk} The stellar disk luminosity density is supposed to be appropriately 
described by an exponential distribution, both horizontally and vertically.
The horizontal scale length is $r_\star = 4$~kpc; the vertical scale height
is $z_\star = 0.35$~kpc; these parameters are fixed to match the observed
values for the old disk population of the Galaxy.
The disk is horizontally and vertically truncated at a distance $r_{max}$ equal
to 6 times the corresponding scale length. 

\underline{Bulge} The luminosity density distribution of this component
is modelled as a Jaffe bulge (see BFG) which reproduces the $r^{1/4}$ profile
characteristic of elliptical and bulge systems. We adopt an effective radius
proportional to the scale length of the disk, $r_e=\eta r_\star$  with the constant
$\eta$ allowed to take the following three values: $\eta=0.1, 0.4, 1.6$.
The bulge is truncated at a distance equal to $r_{max}=5r_e$.

\underline{B/T and Inclination}  To simulate different Hubble types, we
consider different bulge-to-total luminosity ratios. Specifically, we study
the cases B/T=0, 0.1, 0.3, 0.5, 1. which should approximate the 
entire Hubble sequence from Sd to Sa types, respectively. The simulated
galaxy is observed in 9 different inclination bands, centered on the 
angles: $i= 9.31^\circ,  22.9^\circ, 
30^\circ, 40^\circ, 50^\circ, 60^\circ, 70^\circ, 80^\circ, 
90^\circ$ (edge on). 
Edge-on images contains all the photons between 88.5$^\circ$ and 
91.5$^\circ$. The width of each inclination band is set to
cover the same solid angle as the edge-on case (see BFG).

\subsubsection{Dust Distribution}

\underline{Disk} The dust is assumed to be smoothly distributed in
the same plane as the stellar  disk and with a double 
exponential distribution as the stellar one. The horizontal
scale length is the same as for the stars, $r_d = r_\star = 4$~kpc;
the vertical scale height is instead $z_d = \xi z_\star$, with $\xi =
0.4, 1., 2.5$, thus simulating dust that is less or more extended
than the stars, respectively. 

Motivated by the recent suggestions
that dust can be not only vertically but also horizontally more
extended than stars (Ferrara \etal 1991, Ferrara 1997, Davies \etal 1997, 
Xilouris \etal 1997, 1998),
we have also studied a particular case identified by $r_d = 1.5
r_\star$, $\xi=2$ and $\eta = 0.4$.  This case should suitably 
incorporate the recent findings by Davies \etal (1997) on the Galaxy. 

\subsubsection{Model Classification}

The spiral galaxy models can be classified according to 
this system. The first letter is a S for "spiral"; the 2nd and
3th character refer to the value of $\eta$ corresponding to the given 
model. An underscore
introduces the part of the word referring to the dust properties: the adopted
extinction curve (M=Milky Way, S=Small Magellanic Cloud), E
stands for double "Exponential" distribution (one case is labelled ``D" and
refers to the radially extended dust distribution suggested by Davies \etal 1997).
Finally, the remaining two characters give the value of $\xi$. Thus, for example,
the model S01\_ME04 identifies the spiral galaxy model with $\eta=0.1$
with Milky Way-type dust distributed according to a double exponential law, 
whose scale height is $\xi=0.4$ times smaller than the stellar scale height.
For each of these models the entire set of B/T, inclination
and $\tau_V$ values given above has to be simulated. 
A summary of the model properties is given again in Tab. 1.

\subsection{Elliptical Galaxies}

\subsubsection{Stellar Distribution}

We describe the stellar luminosity density distribution of elliptical galaxies 
with a Jaffe bulge. We adopt a fixed  effective radius  $r_e= 4$~kpc  
truncated at $r_{max}=5r_e=20$~kpc. Thus, elliptical galaxies in our scheme
have only the single value B/T=1; also, for models in which the dust
is spherically distributed,  given the complete spherical symmetry of 
the system, it is obviously not necessary to simulate different inclinations.

\subsubsection{Dust Distribution}

For elliptical galaxies we investigate the effects of  the four different dust 
distributions described in the following.

\underline{Model E} Dust is distributed in a double exponential
disk, similarly to the spiral galaxy case, with $r_d=4$~kpc and
$z_d=0.05$~kpc (Cimatti \etal 1997). Inclinations are the same as in that case
as well,  $i=9.31^\circ,  22.9^\circ, 
30^\circ, 40^\circ, 50^\circ, 60^\circ, 70^\circ, 80^\circ, 
90^\circ$ (edge on). 

\underline{Model C} Dust has a constant density inside a sphere 
with the same radius as the bulge, $r_{max} = 5r_e=20$~kpc.

\underline{Model J} This model assumes that the dust is distributed
as the light, \ie with a Jaffe density distribution having the same parameters
as the bulge: $r_e= 4$~kpc, $r_{max}=5r_e=20$~kpc. 
As the density in a Jaffe bulge diverges at $r=0$, we have defined
$\tau_V$ along a line of sight passing through $r=r_e$ for this
model. 

\underline{Model R}  Finally we have considered a model in which
the dust density profile decreases as $r^{-1}$ out to a radius 
$r_{max}=20$~kpc. In analogy with model J, the central divergence of
the adopted profile, forces us to define $\tau_V$ along a line of sight passing 
through a radius $r_o=15.7$~kpc which contains the same dust fraction
(61.6\%) inside $r_e$ in our Jaffe distribution.

\subsubsection{Model Classification}
 
In analogy with spiral galaxies, we introduce a classification scheme 
for elliptical galaxies as well. In this case, the first letter is a E for 
"elliptical". As the stellar component has fixed parameters, no other
information about the light distribution is needed. Again, the  underscore
introduces the part of the string referring to dust properties: the first
character after the underscore denotes the type of extinction curve 
(M=Milky Way, S=Small Magellanic Cloud). The next (and final) character
identifies one of the four studied dust distributions.
Using the nomenclature already introduced in the previous section, we
use "E" for double exponential, "C" for a constant density sphere,
"J" for a Jaffe spherical distribution, and "R" for a $r^{-1}$ spherical
distribution.
Thus, for example, the model E\_ME identifies the elliptical galaxy model with 
with Milky Way-type dust distributed according to a double exponential law,
with parameters defined above.
For E\_ME and E\_SE  models  all the inclination
and $\tau_V$ values given above have to be simulated; as models 
"C", "J", and "R" are spherically symmetric, only runs for the 9 adopted 
values of $\tau_V$ are calculated.
A summary of the model properties is given in Tab. 1.

\section{Attenuation Curves from Monte Carlo Simulations}

In order to calculate the attenuation properties of the various models
of spiral and elliptical galaxies, we have performed a large number
of Monte Carlo simulations. These simulations allow us to treat
the radiative transfer in dusty galaxies (\ie considering the extinction
as the combination of absorption and scattering). The Monte Carlo
code that we use for this study is adapted from the one already
presented in BFG and we defer the interested reader to that paper
for details. The difference consists in the fact that in this work we 
consider a simplified single grain approach, with a Henyey-Greenstein
scattering phase function (Henyey \& Greenstein 1941; see also BFG): 
scattering properties are thus defined by a single value of the albedo 
$\omega$ and of the asymmetry parameter $g$. Values of $\omega$ and $g$ 
are taken from Gordon \etal (1997). The polarization part in the
radiative transfer code has been omitted.

The main output from the Monte Carlo simulations are 
the attenuation curves \att. These are defined as 
\begin{equation}
\label{att}
{\cal A}_\lambda = {{\rm Observed,~dust~extinguished, total~intensity}~I_{obs}(\lambda) 
\over   {\rm  Intrinsic,~unextinguished,~total~intensity}~I_{0}(\lambda)} 
\end{equation}
In general, \att is a function of  $\tau_V$, $i$ and B/T. 
The attenuation curves are calculated for 14 wavelength bands
defined in Gordon \etal  1997: UV1, UV2, UV3, UV4, UV5, UV6, UV7, U, 
B, V, R, I, J, K, with central wavelength $\lambda =$ 1250, 1515, 1775,
1995, 2215, 2480, 2895, 3605, 4413, 5512, 6594, 8059, 12369, 21578 \AA, 
respectively. As an example of the data (available in electronic form 
at the web site http://www.arcetri.astro.it/\~~sbianchi/attenuation.html) 
we show attenuation curves for some cases selected from four representative 
models (summarized in Tab. 1) in Figs. \ref{fig2}-\ref{fig3}.
In addition, Fig. \ref{fig4} shows the wavelength dependence of the fraction  
of the emitted energy absorbed by dust for a spiral galaxy model.
This quantity could be used together with the observed stellar SED
and FIR luminosity, to retrieve the intrinsic {\em unextinguished}
stellar emission. In models with spherical symmetry both for stars and
dust (\eg E\_xC, E\_xJ, E\_xR), the fraction of emitted energy absorbed by 
dust is simply 1-${\cal A}_\lambda$.
These data are also available in electronic form at the above site.

\subsection{Trends in the Attenuation Curves}

We give now a general description of the attenuation curves aimed
at isolating the effects of the various galactic and dust properties
investigated. 

\centerline{\it a) Wavelength dependence}  

The \att curves in all cases show an overall increasing trend   
as a function of the observed wavelength $\lambda$; this closely 
reflects the behaviour of the extinction curves. When the 2200\AA\ 
bump is present in the extinction curve (\ie MW-type),
its footprint is also seen in the attenuation curves, although 
with varying strength, which depends on inclination and optical depth. 
This effect is explained by scattering modulation due to geometry and 
multiple scattering of photons on dust grains, respectively. 

\centerline{\it b) Hubble Type dependence}  

The main difference between the pure disk (B/T=0) and bulge (B/T=1)
cases is that disks are more absorbed at high inclinations (\ie
edge-on) because of the larger optical depth of the dust in the plane
of the galaxy. Bulges, instead, are more attenuated at low $i$, as
in that case basically only the foreground part of the spheroid is 
seen, whereas as $i$ is increased almost the entire bulge comes into
view, provided the dust thickness is not substantial with respect
to the bulge effective radius. This case implies large $\eta$ and low
$\xi$ values.

\centerline{\it c) Dust Optical Depth and Distribution}  

The behaviour of \att with the dust optical depth $\tau_V$ is rather 
straightforward to interpret: more light is absorbed by increasing $\tau_V$
although in some cases saturation at $\tau_V \simgt 5$ occurs, particularly
in UV/optical bands. 
The effect of increasing $\eta$, \ie more extended bulges, typically
causes, for a given value of $\tau_V$ and $i$, the galaxy to be less
attenuated. This is due to the fact that a larger fraction
of the bulge emitted light is outside the dust obscuring layer. 
The amount of attenuation decreases as $i$ is increased, 
both for disk- and bulge-dominated galaxies.
However, there are exceptions to this rule for early Hubble
types, where the bulge is particularly pronounced (\eg S16\_ME04
model) or small compared to the dust scale height (\eg S01\_ME25): 
in the first case more edge-on galaxies allow for a larger fraction 
of the galactic bulge light to escape freely; in the second case, inclination
effects are basically negligible as the bulge is only seen through
the dust layer at any inclination to the line of sight. 
 
The thickness of the dust layer, described by the parameter $\xi$
profoundly affects the observable spectral energy distribution
of the galaxy. An increase of a factor 2.5 in the value of $\xi$
typically causes a drop in \att by $\approx 0.1$ at UV/optical
wavelengths, whereas the change in the near IR (J and K bands)
is much smaller. For large values of $\xi= 2.5$, and of the optical  
depth, only a fraction $\simlt 1$\% of the emitted photons can escape
from the galaxy (see model S01\_ME25). This  effect is particularly evident
for small bulges, which therefore can be almost totally obscured in 
the UV/optical bands even at relatively low $\tau_V$, a fact that
can have important consequences, as mentioned in \S~1.

The class of models implementing horizontally extended dust distributions
(S04\_MD20 and S04\_SD20) are much more extinguished than other models
with comparable vertical dust scale height. Disk-dominated galaxies
are obviously more sensitive to this different dust distribution than
bulge-dominated objects, as stars are intermixed with sufficiently 
optically thick dust in a larger area of the disk. 

\centerline{\it d) Extinction Curves }  

Different extinction curves do not seem to produce relevant 
qualitative differences, apart from the presence/absence of
the 2200\AA\ bump in the attenuation curves. Quantitative 
differences are restricted to a few percent level. 

\centerline{\it e) Elliptical Galaxies }  

Finally, among the models investigated for elliptical galaxies
the exponential dust distribution (model E\_ME) produces the smallest
attenuation factors (in this case the dust disk scale height is
relatively thin, $z_d = 50$~pc). Among the spherically symmetric
models, the constant (E\_MC), and $1/r$ (E\_MR) distributions 
give similar results being able to efficiently attenuate the most 
luminous central regions of the bulge. For the Jaffe distribution, in
which the dust is well intermixed with the stars, the attenuation is
considerably reduced with respect to the other two above cases. 

\section{An Application to Local Spirals}

The models presented above are particularly suitable to 
study the effects of dust in distant galaxies, 
for which only integral properties as, for
example, the total apparent magnitude in a given band is known.  

Nevertheless, testing our predictions against
local observational data allows to calibrate
our models to the nearby universe. In addition, the 
derived galactic properties (\eg dust optical depth,
geometry, etc)  might serve as a guide when exploring their 
counterparts in the young universe. 

As an example, in the following we study the dependence of the
total apparent magnitude $m$ of a galaxy in a given band on the
inclination angle $i$. Obviously, in the absence of dust,
there would be no reason to expect a correlation between these 
two quantities in a sample of galaxies of the same mass, as from
the Tully-Fisher relation it follows that they should consequently have the
same luminosity. The inclination dependence of apparent magnitudes
is instead now well established (de Vaucouleurs \etal 1991, 
Giovanelli \etal 1994, Peletier \& Willner 1992, Peletier \etal 1994,
Evans 1994, Berstein \etal 1994) and thought to depend on dust obscuring
effects. This internal attenuation results in a magnitude correction $\Delta m$
which is inclination-dependent, in the sense that more edge-on galaxies
appear fainter. In Fig. \ref{fig5} we compare the predicted values $\Delta m = 
-2.5\log$\att with the fit to the observational values derived by
de Vaucouleurs \etal (1991) [$\Delta m= 1.45 \log (1/\cos~i)$] in the
B-band, and by Bernstein \etal (1994) [$\Delta m= 1.4 (1-\cos~i)$] in the
I-band. As both samples mostly contain Sb-Sd galaxies, we choose B/T=0.1
models, MW-type extinction, and use the value $\eta=0.4$
appropriate for the Milky Way (see BFG); this defines the class of 
models S04\_MExx. For each band we compare the
data with $\Delta m$ curves as a function of $1-\cos~i$ for three values 
of $\xi=0.4, 1.0, 2.5$ and 4 values of $\tau_V= 0.5, 1, 5, 10$. The 
abscissa limits are dictated by the availability of data for different
inclinations. The conclusion from the inspection of Fig. \ref{fig5} is rather    
straightforward: only models which assume a central $\tau_V \sim 5-10$ 
can simultaneously fit the data in both bands; also, models in which 
the dust thickness is larger (higher $\xi$) seem to perform better. 
Note that the vertical offset between the data points and the curves (they
have been normalized at the same value at $i=60^\circ$ or  $(1-\cos~i)=0.5$) is 
arbitrary as the dust-free luminosity cannot be experimentally determined. 
This simple example shows how our models are able to satisfactorily reproduce
the high quality data available locally; therefore, unless the properties
of dust and galaxies are considerably different in the young universe, the 
hope is that they could serve to robustly model the effects of dust at
high redshift as well. 

\section{Clumping effects}

As a final remark, we like to recall that our models do not include yet 
the possible existence of a clumpy dust component in addition to the 
smoothly distributed one considered. The extension of the present work
to such a case is already in progress; here we would like to discuss briefly
the consequences of our smooth distribution assumption deduced from these 
preliminary results. 

High redshift galaxies often show rather irregular morphologies 
suggestive of inhomogeneous dust and light distribution 
(Abraham \etal 1996), possibly produced by the higher merging/interaction
rates at those epochs. On the other hand, Giavalisco \etal 1996 found
that UV dropouts at $z>3$ either show a core similar
to present day spheroids or are described by exponential luminosity profiles,  
thus providing a solid observational basis for our assumptions.
 
As a general rule, clumpiness typically results in a lower effective 
dust optical depth and may affect the wavelength dependence of the attenuation
curves. Estimating by how much our models might be overestimating 
the attenuation is not trivial as, if we allow for the dust 
to partly reside in clumps, at least three additional parameters should
be introduced. These are: the clump/interclump dust density contrast, 
the size of the clumps and the filling factor of clumps. The above parameters
are only loosely constrained by current observations. The mere schematization
of the dust distribution as a two-component medium made of clumps and
interclump medium escapes a clear physical interpretation, as several possibilities
arise from the complexity of the ISM: (i) can the dust clumps be identified
with molecular clouds ?  In this case they should be distributed in the
galaxy following the molecular hydrogen abundance, rather than randomly; 
(ii) are they the so-called "clumpuscules" proposed by several authors
as dark matter candidates ? if so, their sizes and masses 
should be pretty small (30 AU and Jupiter mass, respectively; Pfenninger
\& Combes 1994); (iii) do they arise in fractal-like density fluctuations in
the ISM ? A spectrum of sizes should be then expected and the modelling
would become computationally quite expensive; finally, (iv) are they a 
reflection of transient coherent features as dust lanes or produced by
dynamical effects as galaxy interactions, merging, stripping or 
spiral density waves ? It is clear that the very nature of these clumps 
strongly affects that way in which we include clumpiness in the models or we 
interpret its observational consequences. A lot of effort to understand
these issues from the experimental point of view is necessary to make
a progress. 
Even if the value of the free parameters would be perfectly known, 
the only way to assess their effects would 
be to perform Monte Carlo simulations of the type presented in this paper.
Our preliminary runs including a clumpy component modelled as spherical clouds
of different size and number, suggest that the attenuation
curves start to become insensitive to presence of inhomogeneities above
a critical inclination angle $i_{c}(f)$ which depends on the value of the 
clump filling factor $f$.  
Already for $f\approx 0.2$ the critical inclination approaches values as low
as $i_c = 40^\circ$; for higher values of $f$ the differences in the
attenuation curves are almost unnoticed. This conclusion, although 
preliminary, seems to be in agreement with the findings of recent/ongoing 
work by different groups (Byun \& Lee 1997, Kuchinski \etal 1998).
If confirmed, these arguments imply that the effect of clumping is
much less dramatic than usually believed, although further study - both
theoretical and observational - is clearly needed.

\section{Attachments}

The electronic source files containing the attenuation (of the type
shown in Figs. \ref{fig2}-\ref{fig3}) and energy absorption (Fig. \ref{fig4})
curves for the entire set of models described here and 
summarized in Tab. 1, can be retrieved from the web
site http://www.arcetri.astro.it/\~~sbianchi/attenuation.html. 

\vskip 1truecm

We thank C. Lacey for stimulating discussions and for suggesting
the comparison with local galaxies and the referee, J. Mathis, for
very insightful and useful comments. 

\newpage
\section{Appendix A}

In the following we give the formulae allowing to convert the value
of the optical thickness $\tau_V$ into a total galactic dust mass
for the four dust distributions considered in this paper,
with the assumptions for the average grain radius, $a$, V-band extinction
coefficient, $Q$, and grain density, $\delta$ given below.

$$a=0.1 \mu{\rm~m,}{\rm average~grain~radius}$$
$$Q=1.5, {\rm~V-band~extinction~coefficient}$$
$$\delta=2 {\rm~g~cm,}^{-3}{\rm grain~density}$$

\underline{\bf exponential~disk}

$$M=\tau_V\;\alpha_d^2\;\frac{8\pi}{3}\frac{a\delta}{Q}\;\left[1-(n+1)e^{-n}\right]
\;\;\;\; (n=6 {\rm ~truncation~factor})$$
$$M\left[M_\odot\right]=5.3\times 10^5\;\tau_V\;(\alpha_d\left[{\rm kpc}\right]
)^2$$
$$M\left[M_\odot\right]=8.4\times 10^6 \;\tau_V\;\;\;\;\; \alpha_d=4{\rm~kpc}$$
$$M\left[M_\odot\right]=1.9\times 10^7 \;\tau_V\;\;\;\;\; \alpha_d=6{\rm~kpc}$$

\underline{\bf constant sphere}

$$M=\tau_V\;r_{max}^2\;\frac{8\pi}{9}\frac{a\delta}{Q}$$
$$M\left[M_\odot\right]=1.8\times 10^5\;\tau_V\;(r_{max}\left[{\rm kpc}\right])
^2$$
$$M\left[M_\odot\right]=7.1\times 10^7\;\tau_V\;\;\;\;\;\; r_{max}=20{\rm kpc}
$$

\underline{\bf Jaffe sphere}

$$M=\tau_V\;r_0^2\;\frac{8\pi}{3}\frac{a\delta}{Q}\;\left[
\frac{\frac{n}{n+1}}
{{\rm asin}(1)-{\rm asin}(1/n)-\frac{\sqrt{n^2-1}(5+4n)}{3(1+n)^2}}
\right]$$
$$r_0=4.64 {\rm~kpc}\;\;{\rm~(corresponding~to~}r_e=4 {\rm~kpc})\;\;\;\;r_{max}
=20 {\rm~kpc}\;\;\;\;n=r_{max}/r_0$$
$$M\left[M_\odot\right]=1.8\times 10^6 \;\tau_V\;r_0^2$$
$$M\left[M_\odot\right]=4.0\times 10^7 \;\tau_V$$

\underline{\bf $1/r$ sphere}

$$M=\tau_V\;r_0^2\;\frac{4\pi}{3}\frac{a\delta}{Q}\;\left[
\frac{n^2}{ {\rm ln}\left(n+\sqrt{n^2-1}\right)}
\right]$$
$$r_0=15.7 {\rm kpc}\;\;\;\;r_{max}=20 {\rm kpc}\;\;\;\;n=r_{max}/r_0$$
$$M\left[M_\odot\right]=6.0\times 10^5 \;\tau_V\;r_0^2$$
$$M\left[M_\odot\right]=1.5\times 10^8 \;\tau_V$$

\newpage
\parindent 0.0cm

\begin{figure}
\centerline{\psfig{figure=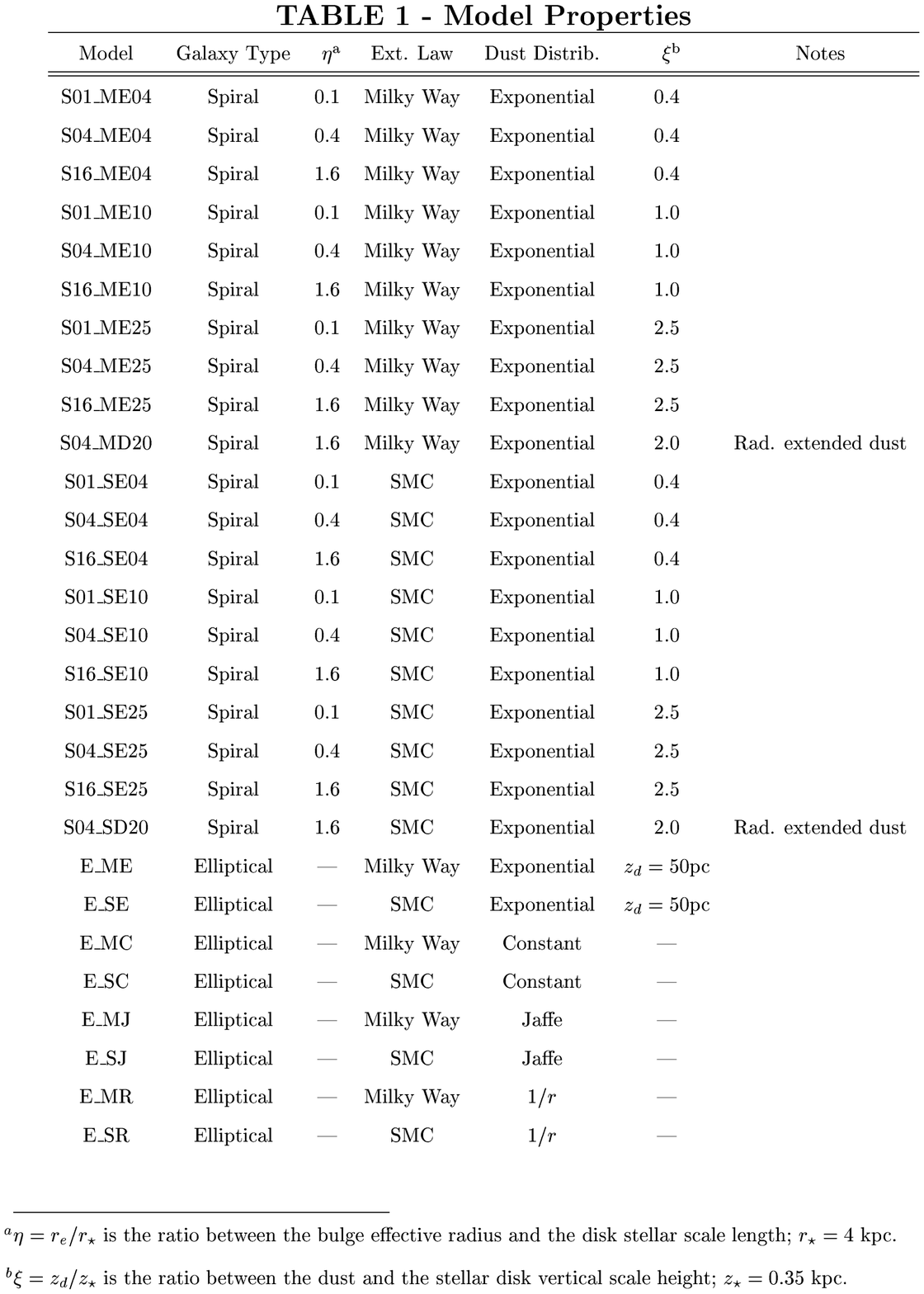,height=26cm}}
\end{figure}
\begin{figure}
\centerline{\psfig{figure=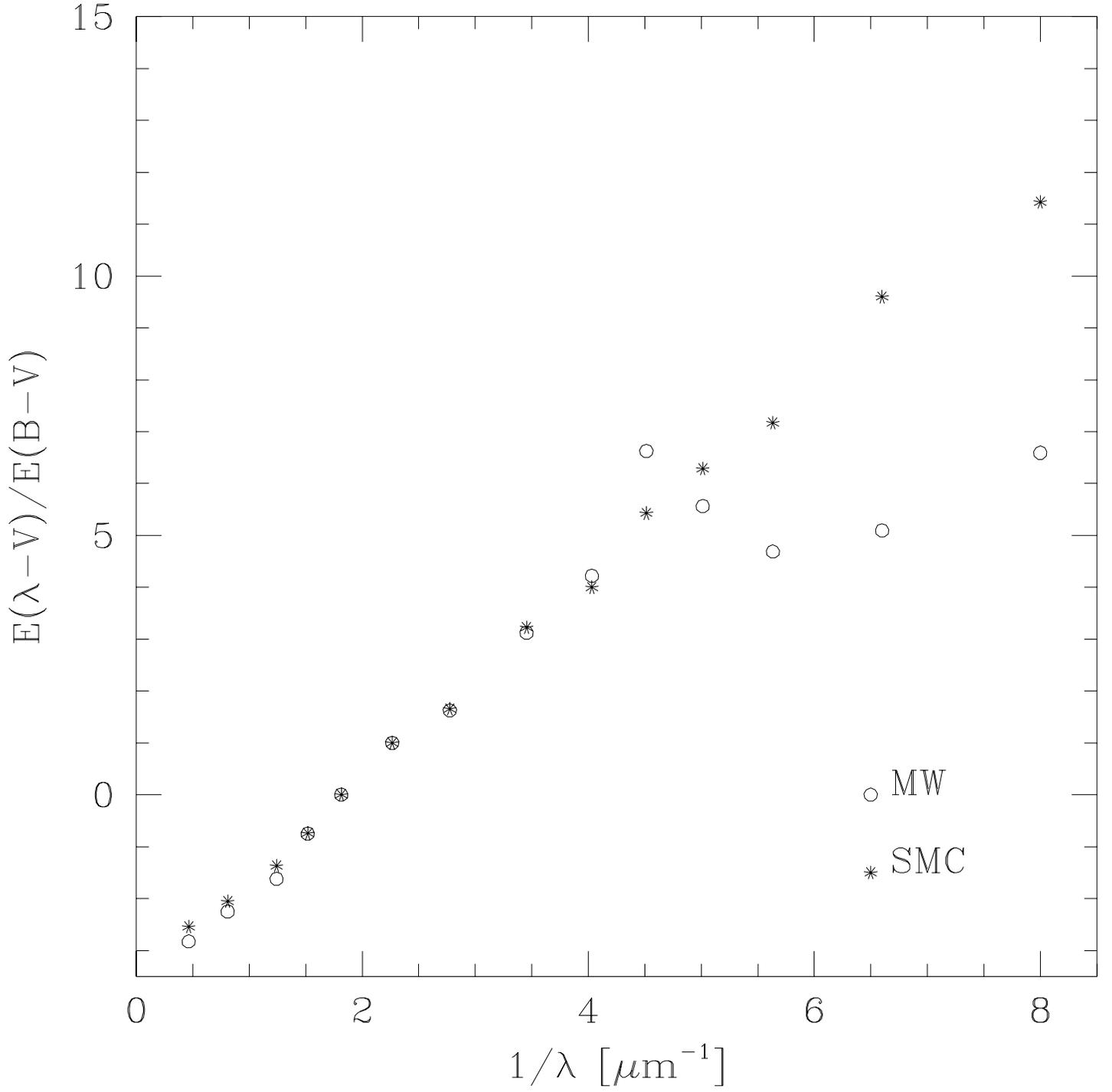}}
\caption{\label{fig1}{\small Adopted extinction curves for the Milky Way (open
circles)  and the Small Magellanic Cloud (stars).}}
\end{figure}

\begin{figure}
\centerline{\psfig{figure=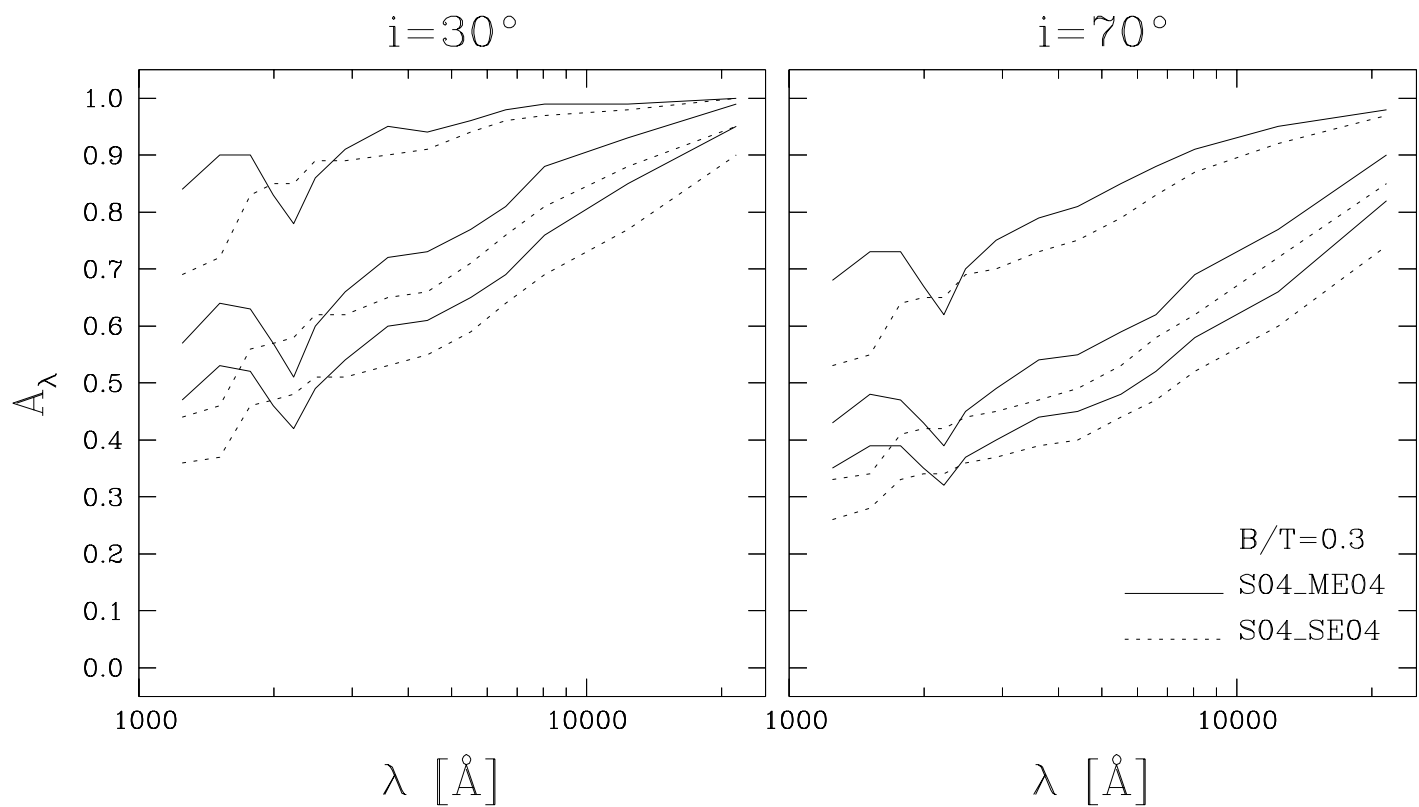}}
\caption{\label{fig2}{\small        Attenuation curves for the model S04\_ME04 ({\it solid} curves)
and S04\_SE04 ({\it dotted}) as a function of wavelength; model parameters are defined 
in Tab. 1. The curves are shown for bulge-to-total ratio B/T=0.3,  two  
inclination angles  $i=30^\circ, 70^\circ$, and optical depths $\tau_V=1, 5, 10$
from the uppermost to the lowermost curve in each panel.}}
\end{figure}

\begin{figure}
\centerline{\psfig{figure=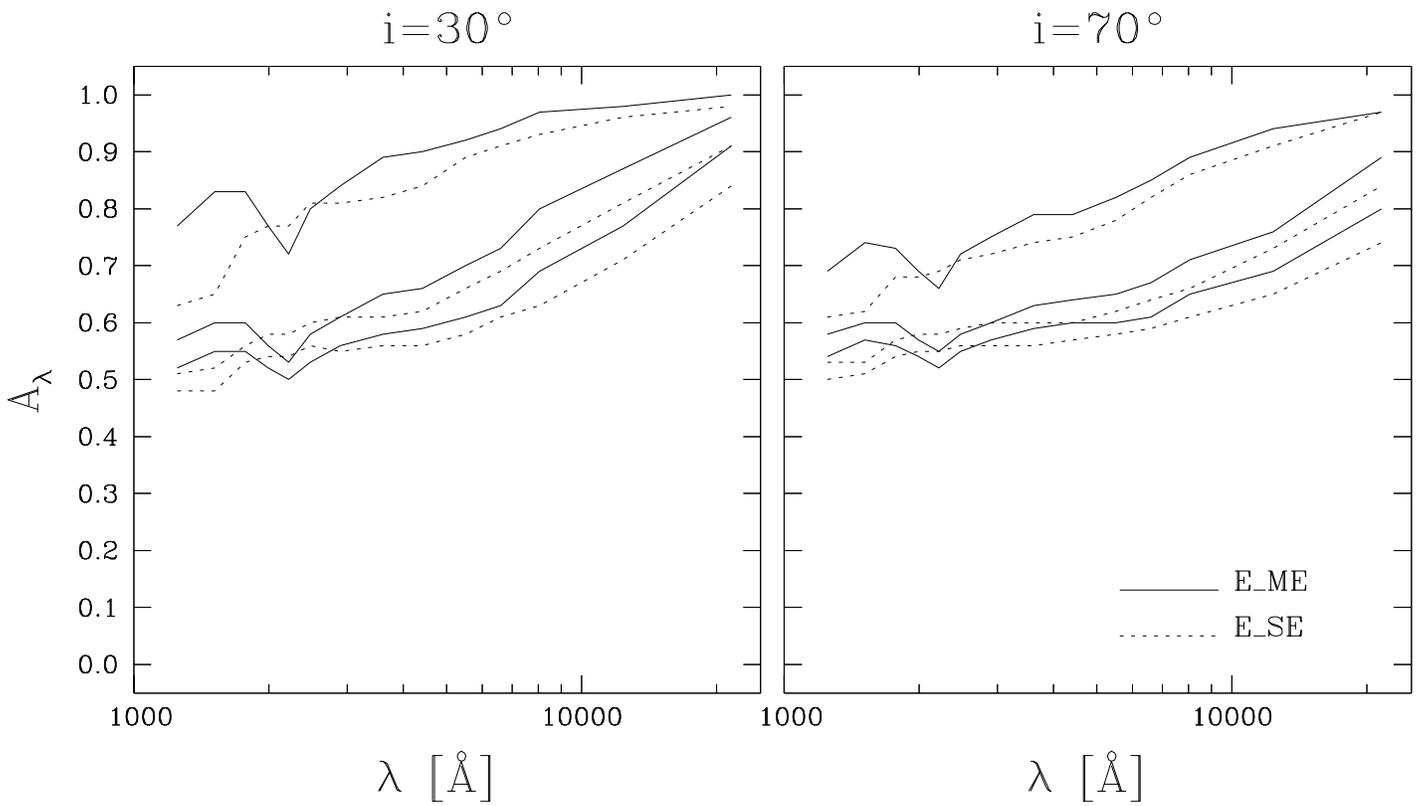}}
\caption{\label{fig3}{\small   Same as Fig. \ref{fig2} for the models 
E\_ME and E\_SE; model parameters are defined in Tab. 1.}}
\end{figure}

\begin{figure}
\centerline{\psfig{figure=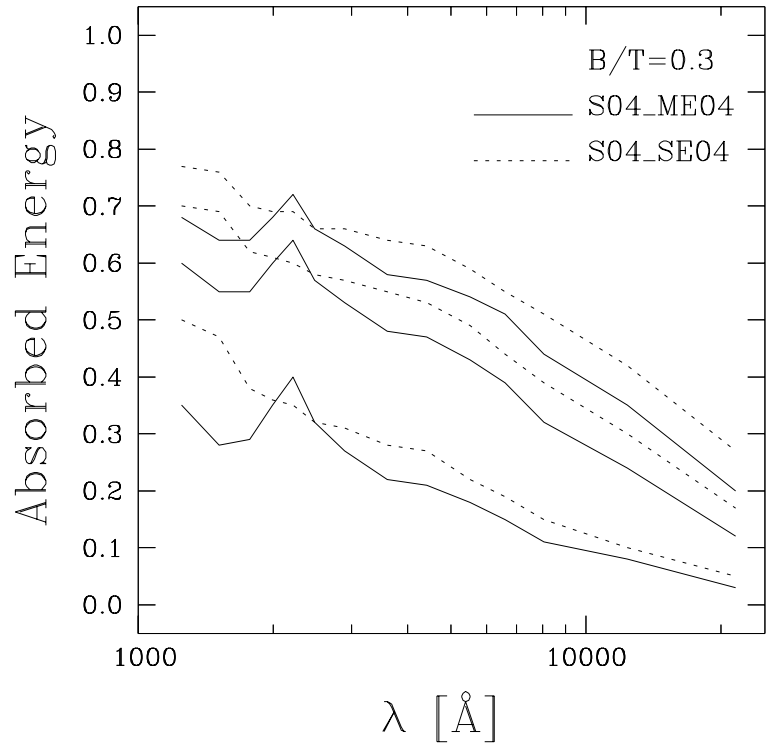}}
\caption{\label{fig4}{\small Fraction of the emitted energy absorbed by dust 
as a function of wavelength for the models S04\_ME04 ({\it solid} curves)
and S04\_SE04 ({\it dotted}) as a function of wavelength; model parameters are defined
in Tab. 1. The curves are shown for bulge-to-total ratio B/T=0.3 
and optical depths $\tau_V=1, 5, 10$ from the lowermost to the uppermost curve in each panel.}}
\end{figure}

\begin{figure}
\centerline{\psfig{figure=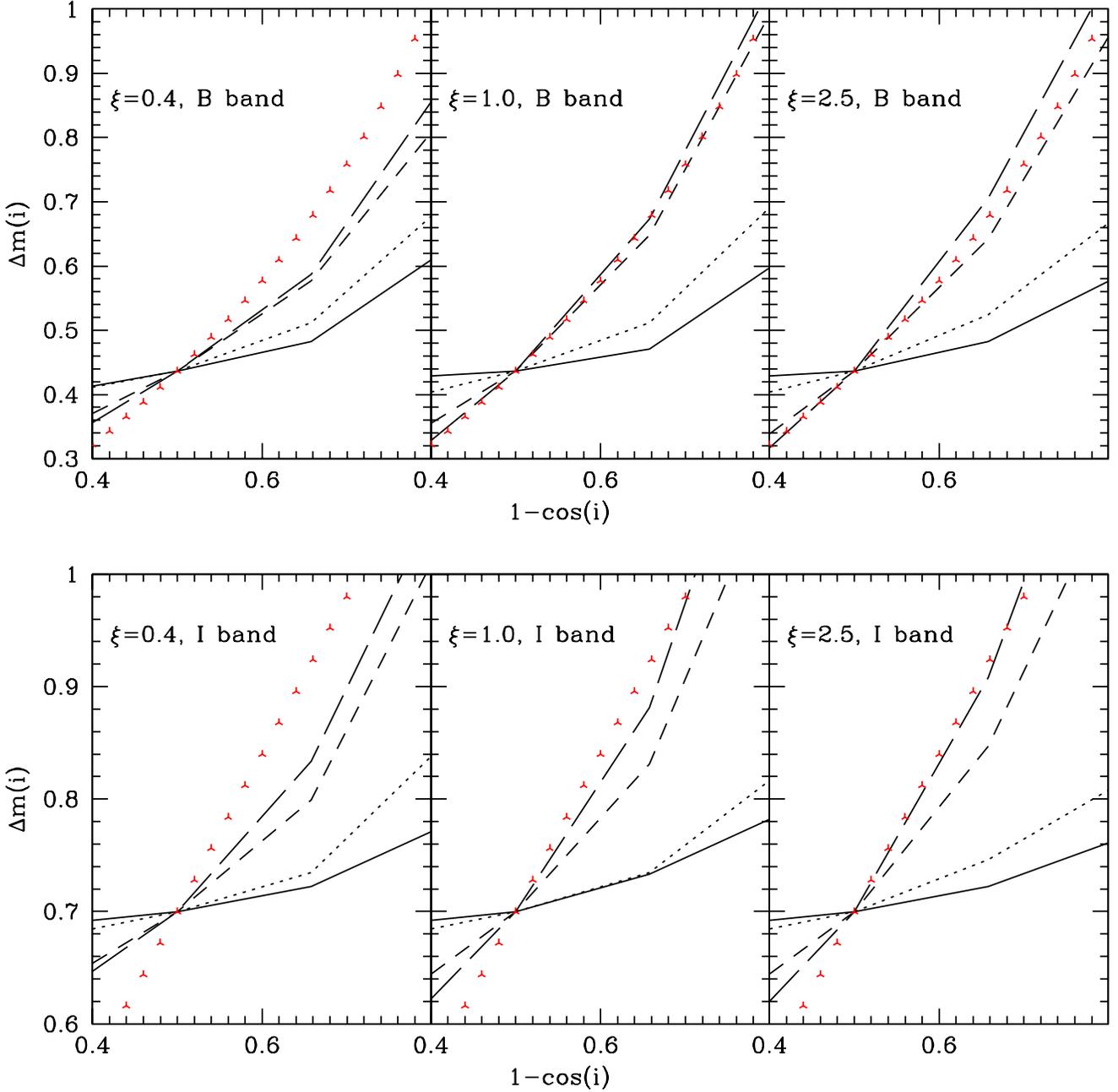}}
\caption{\label{fig5} {\small Dependence of the internal extinction correction $\Delta m$
on $1-\cos(i)$ for the B-band (upper three panels) and I-band (lower
panels). Points are observational data from de Vaucouleurs \etal (1991)
(B-band) and Bernstein \etal (1994) (I-band); note that the vertical
offset of the data is arbitrary. Curves are from models S04\_MExx for B/T=0.1, where
xx=04,10,25 (or $\xi =0.4, 1.0, 2.5$); each line refers to a different
value of $\tau_V$=0.5 (solid), 1 (dotted), 5 (short-dashed), 10 (long-dashed).}}
\end{figure}

\end{document}